\documentclass[fleqn,usenatbib]{mnras}

\usepackage{mathptmx}

\usepackage[T1]{fontenc}

\DeclareRobustCommand{\VAN}[3]{#2}
\let\VANthebibliography\thebibliography
\def\thebibliography{\DeclareRobustCommand{\VAN}[3]{##3}\VANthebibliography}

\usepackage{graphicx}	
\usepackage{amsmath}	
\usepackage{amssymb}	
\usepackage{threeparttable} 
\usepackage{float} 
\usepackage[dvipsnames]{xcolor} 
\usepackage{soul} 
\usepackage{subcaption} 

\title{Stellar clustering and the kinematics of stars around Collinder 121 using \textit{Gaia} DR3}

\author[G. D. Fleming]{
Graham D. Fleming,$^{1}$\thanks{E-mail: gdfleming@uclan.ac.uk}
Jason M. Kirk,$^{1}$
Derek Ward-Thompson$^{1}$
\\
$^{1}$Jeremiah Horrocks Institute, University of Central Lancashire, Preston, PR1 2HE, UK\\
}

\date{Accepted XXX. Received YYY; in original form ZZZ}

\pubyear{2022}

\begin{document}
\label{firstpage}
\pagerange{\pageref{firstpage}--\pageref{lastpage}}
\maketitle

\begin{abstract}

We study the region around Collinder~121 (Cr~121) using newly available 6-dimensional data from the \textit{Gaia}~DR3 catalogue. Situated in the third quadrant, near the galactic plane, Collinder~121 lies in the region of Canis Major centred around (\textit{l,~b})$\approx$(236$^{\circ}$,~-10$^{\circ}$). Previous studies have suggested that the stellar associations in this region comprise an OB association (CMa~OB2) lying at about 740~pc with a more distant open cluster (Cr~121) at approximately 1170~pc. Despite these studies, the precise nature of Collinder~121 remains uncertain. This study investigates the region bounded by the box (\textit{l,~b}) = (225$^{\circ}$ to 245$^{\circ}$,~0.00$^{\circ}$ to -20.00$^{\circ}$) to a depth of 700~pc from 500 to 1200~pc which fully encompasses the region discussed in the literature. Using \textit{Gaia}~DR3 data, we do not find associations at the distances given in the literature. Instead, using the HDBSCAN machine learning algorithm, we find a major association of OB stars centred around 803~pc. Within this association we find four smaller subgroups that may be indicative of a larger association and which are located at a mean distance of 827~pc. Proper motion studies find coherence between these four subgroups and show a distinctive east to west increase in the size of the velocity vectors which supports contemporary studies that show similar trends in OB populations in Cygnus and within the Carina spiral Arm. Therefore, we hypothesize that Cr~121 and CMa~OB2 are the same cluster, consistent with the 1977 study by Hoogerwerf.

\end{abstract}





\section{Introduction} 
\label{Intro}

\begin{figure} 
    \includegraphics[width=0.95\columnwidth]{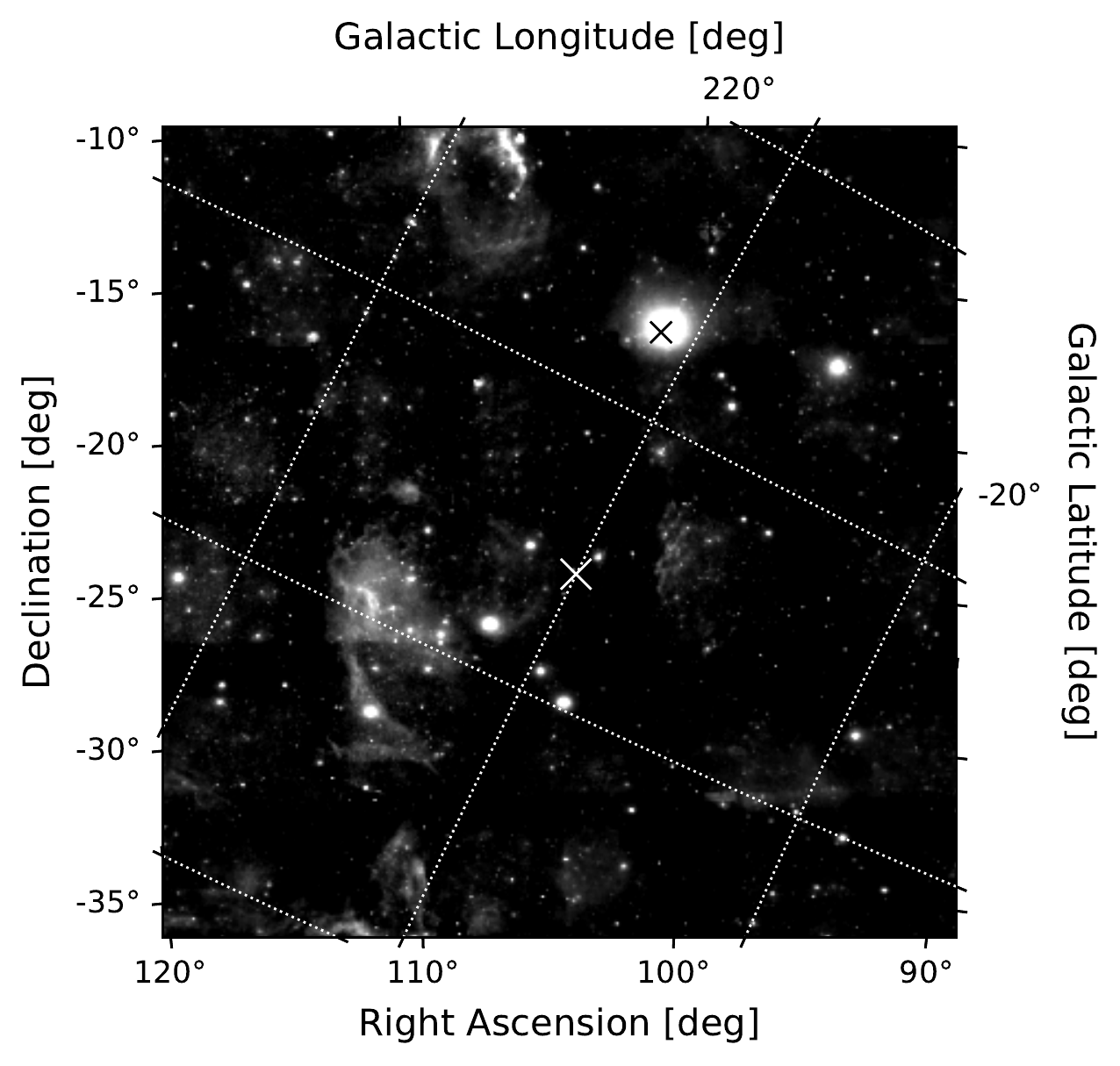}
    \caption[Location of Collinder~121.]{The region of Collinder~121 in relation to the constellation of Canis Major. The image is bounded by the limits defined in our \textit{Gaia} EDR3 query. The white cross is centred on (\textit{l,~b}) = (235$^{\circ}$,~-10$^{\circ}$), the centre of Collinder 121 \citep{wu2009orbits}. For reference, the black cross denotes the position of the A0 star $\alpha$CMa (Sirius) \citep{van2007validation}.\\
    \small DSS Image from: Aladin Sky Atlas, \href{https://aladin.u-strasbg.fr/}{https://aladin.u-strasbg.fr/}. \citep{bonnarel2000aladin}, (Modified here).}
    \label{fig:CanisMajor}
\end{figure}

Published in 1931 by the Swedish astronomer Per Collinder following his studies at Lund University, the Collinder catalogue details the structural and spatial properties of 471 open galactic clusters. The catalogue was published as an appendix to Collinder's paper \citep{collinder1931structural}. Objects in the catalogue are denoted by their catalogue number and cover the entire galactic region between declinations +85 degrees and -79 degrees. Previous studies of the region around Collinder~121 (Figure~\ref{fig:CanisMajor}) have sought to determine possible cluster membership using common motion studies. Photometric and astrometric studies \citep[e.g.][]{de1999hipparcos, burningham2003nature, perez2005search, kaltcheva2007structure} have investigated discrete areas within the overall region but have not provided a consistent morphology.

A stellar group of 20 main-sequence B-stars in this region was designated as catalogue number 121. Spanning an area of 50$\times$35~arcmin located at (\textit{l,~b}) $\sim$ (202.5$^{\circ}$,~-9.6$^{\circ}$) (epoch 1900)\footnote{(\textit{l,~b}) $\sim$ (203.1$^{\circ}$,~-8.5$^{\circ}$) (epoch 2000)}, the grouping was placed, through the use of an unpublished card system compiled in 1919 by Lundmark \citep[see e.g.][]{teerikorpi1989lundmark} and observations of stellar magnitudes, at a distance of 1910~light-years (586~pc). \cite{feinstein1967collinder} revisited the region and concluded that the Cr~121 open cluster consisted of a red supergiant (HD~50877~/~omi01~CMa) and a collection of main-sequence B-stars lying at 630~pc within a 1$^{\circ}$ circle in Canis~Major. In addition, \cite{feinstein1967collinder} noted the possibility of other members extending within a 10$^{\circ}$ circular region and suggested that Cr~121 resembled a double cluster.

In a study of NGC~2287 and the Pleiades group, \cite{eggen1974ngc} suggested that an association of stars surrounding Collinder~121 might be moving with the Pleiades group, but noted that this possible membership was based on poor proper motion data and that the \textit{UVW} vectors were calculated on the strength of the radial velocity data alone. In his later study, \cite{eggen1981region} identified two separate regions of stellar density, one an association at 700-900~pc, the other an open cluster-like group at 1000-1100~pc. Both groups were described as being indistinguishable, other than being \textit{"separated in the line of sight by nearly 500 pc with few, if any, luminous stars between.} \citep{eggen1981region}.

\cite{hoogerwerf1997new}, using \textit{Hipparcos} data derived a distance of 546$\pm$30~pc for Collinder~121, concluding that, as there was no clear difference in the parallax data for the two groups proposed by \cite{eggen1981region}, there was no division in the stellar groups.

As part of their comprehensive study of OB associations within 1~kpc of the Sun, \cite{de1999hipparcos} investigated the astrometric membership of Collinder~121 using data from \textit{Hipparcos}. Using a combination of the convergent point method \citep{bruijne1999refurbished} and the 'Spaghetti method' of \cite{hoogerwerf1999identification}, \cite{de1999hipparcos} identified a grouping of 103 stars in an area of 100~pc$\times$30~pc and derived a mean distance of 592$\pm$28~pc for the group. The inclusion of 29~CMa as an O7 class star \citep{sota2014galactic} and numerous early-type B stars attested to the young~($\sim$5~Myr) age of this group. The study found two subgroups centered on (\textit{l,~b})$\approx$(233$^{\circ}$,~-9$^{\circ}$) and (238$^{\circ}$,~-9$^{\circ}$) with the suggestion of a possible third group lying at (\textit{l,~b})$\approx$(243$^{\circ}$,~-9$^{\circ}$), all at a distance of $\sim$500~pc.

A photometric study of all stars earlier than spectral type A0 in a field of 5$^{\circ}$ radius centred on the classical centre of the Cr~121 open cluster at (\textit{l,~b})$\sim$(235$^{\circ}$,~-10$^{\circ}$) was conducted by \cite{kaltcheva2000region}. A compact group of stars at 1085$\pm$41~pc was identified, confirming the \cite{collinder1931structural} identification of the Cr~121 open cluster. \cite{kaltcheva2000region} described this group as being \textit{"clearly distinguishable"} from the surrounding field stars located at 660 to 730~pc.

In their later discussion of the region, \cite{kaltcheva2007structure} re-evaluated the parallax data presented in the \textit{Hipparcos} catalogue in light of \textit{ubvy}$\beta$ photometry from other studies. Systematic errors in the \textit{Hipparcos} data were identified and corrected using a method proposed by \cite{makarov2002computing} in their study of the Pleiades open cluster. The re-evaluated findings found a moving group with considerable depth and a complex morphology located at $\sim$740~pc with the cluster being confirmed at a distance of $\sim$1085~pc.

The photometric study conducted by \cite{burningham2003nature} used \textit{XMM-Newton} and \textit{ROSAT} data to investigate the X-ray bright low-mass pre-main sequence population towards the Cr~121 open cluster originally proposed by Collinder \citep{collinder1931structural}. Using isochrones based on solar metallicity \citep[see][]{jeffries2001photometry} \cite{burningham2003nature} suggest that the low-mass PMS stars in their study are more closely linked to a young cluster at 1050~pc and that the group studied by de~Zeeuw is more likely a foreground association as suggested by \cite{eggen1981region}.

Using photometric data gathered from a number of catalogues \cite{perez2005search} present a study of brown dwarfs in the region of the Collinder~121 open cluster. Utilizing 2MASS photometry, they re-analysed the results of \cite{burningham2003nature} and suggested that the results from both studies are consistent with a 11.3~Myr cluster at a distance of 471~pc.

As described, previous studies of associations within Collinder~121 have investigated different types of object in different areas within the region. To summarise the findings of these studies, Table~\ref{tab:Collinderstudies} presents the mean distance and regional limits of the associations identified in each study. It can be seen that there is no consensus in the literature.

\begin{table} 
\centering
    \caption{Parameters of Collinder~121 from the literature. For the purposes of this study we discount the Canis~Major~OB1 (CMa~OB1) stellar association as being outside our study region.}
    \label{tab:Collinderstudies}
    \begin{threeparttable}
    \begin{tabular}{l@{\hspace{1.0\tabcolsep}}l@{\hspace{1.0\tabcolsep}}c@{\hspace{1.0\tabcolsep}}c@{\hspace{1.0\tabcolsep}}c} 
    \hline
    Reference & Assoc. & $D_{min}$ & $D_{mean}$ & $D_{max}$\\
    {} & {} & [pc] & [pc] & [pc]\\
    \hline
    \cite{collinder1931structural}\tnote{a} & Cr~121 & \dots & 590 & \dots\\
    \cite{feinstein1967collinder} & Cr~121 & \dots & 630 & \dots\\
    \cite{eggen1981region} & Cr~121 & \dots & 1170 & \dots\\
    {} & CMa OB2 & \dots & 740 & \dots\\
    \cite{hoogerwerf1997new} &  CMa OB2 & 516 & \dots & 576\\
    \cite{de1999hipparcos} & CMa OB2 & 564 & \dots & 620\\
    \cite{kaltcheva2000region} & Cr~121 & 1044 & \dots & 1126\\
    \cite{burningham2003nature} & Cr~121 & \dots & 1050 & \dots\\
    \cite{perez2005search}\tnote{a} & Cr~121 & \dots & 471 & \dots\\
    \hline
    Mean literature values & & & & \\
    \hspace{5mm} Open Cluster (Cr~121) & & 1044 & 1109 & 1170\\
    \hspace{5mm} OB Association (CMa OB2) & & 516 & 628 & 740\\
\hline
\end{tabular}
    \begin{tablenotes}
    \item[a]\footnotesize{Values not used in determining given mean values.}
   \end{tablenotes}
 \end{threeparttable}
\end{table}

In the following investigation we use the results of our \textit{Gaia}~DR3 query, described in Section~\ref{Data}, to perform a 6-dimensional analysis of Collinder~121 in order to clarify the nature of the stellar associations in the region. Section~\ref{Gaia} provides a brief overview of the \textit{Gaia} mission and some of the relevant issues regarding DR3 and its early release. The acquisition of data is discussed in Section~\ref{Data}. Section~\ref{Discussion} presents a discussion of our data which is briefly summarised in Section~\ref{Summary}. A compendium of OB sources belonging to coherent subgroups identified in this study are presented in Appendix~\ref{DR3 4subgroups}.

\section{\textit{G\lowercase{aia}}} 
\label{Gaia}

The European Space Agency \textit{Gaia} astrometric space observatory \citep{lindegren1996gaia} was launched in December 2013. \textit{Gaia} is designed to measure the parallax, positions and proper motions of stars with the task of producing a three-dimensional map of about one billion stars as they orbit the centre of our Galaxy.  \textit{Gaia} is not designed to measure distances directly, but they can be inferred through the determination of stellar parallax. Data from the mission are stored in the \textit{Gaia} Archive\footnote{\href{http://gea.esac.esa.int/archive/}{http://gea.esac.esa.int/archive/}}, a relational database that can be accessed through an interactive user interface and interrogated using conditional ADQL queries \citep[e.g.][]{possel2019beginner}.

\textit{Gaia}'s~Data~Release~3~(DR3) \citep{vallenari2022gaia} was released on 13th June 2022 and followed on from the previous Early~Data~Release~3~(EDR3) \citep{smart2020gaia} which was released on the 3rd~December~2020.

The DR3 catalogue provides positions on the sky, parallax, and proper motions for around 1.46~billion (1.46 x 10$^{9}$) sources, with a limiting G-band (330 to 1050~nm) magnitude of about G~$\approx$~21 and a bright limit of about G~$\approx$~3. In addition, radial velocities for 33~million sources and spectral classifications for 217~million stars are available in this release for the first time.

\textit{Gaia}'s G-band photometric uncertainties range from $\sim${0.3}~mmag for G~<13~to~6~mmag for stars brighter than 20~mag. Median parallax errors are 0.02~to~0.03~mas for sources brighter than G~<~15~mag, increasing to 1.3~mas for the fainter (G~$\approx$~21~mag) sources. This fainter limit equates to a mean error of of 17~pc at the distances discussed in Section~\ref{FindOB}. The uncertainties in proper motions for these sources are 0.02~–~0.03~mas~yr$^{-1}$ and 1.4~mas~yr$^{-1}$ respectively.

\cite{katz2022gaia} report that the median formal precision of the \textit{Gaia} DR3 radial velocities is of the order of 1.3~km$^{-1}$ at $G_{RVS}$~=~12 and 6.4~km$^{-1}$ at $G_{RVS}$~=~14 mag and is in satisfactory agreement with other surveys such as  APOGEE, GALAH, GES and RAVE.

As such, each successive data release radically increases the accuracy to which open clusters and stellar associations can be measured.

\textit{Gaia}~DR3 is based on data collected between 25th~July~2014 and 28th~May~2017, covering a period of 34~months of data collection. {Gaia} DR3 astrometry uses the ICRS reference system and provides stellar coordinates valid for epoch J2016.5 (roughly mid-2016, where J stands for Julian year).

\subsection{Data Acquisition} 
\label{Data}

An asynchronous ADQL query was submitted to the \textit{Gaia} DR3 archive covering the region (\textit{ra,~dec})~(89.00$^{\circ}$ to 120.00$^{\circ}$,~-10.00$^{\circ}$ to -37.00$^{\circ}$). This region is therefore centred on (\textit{ra,~dec})~(104.5$^{\circ}$, -23.5$^{\circ}$) or, in galactic coordinates (\textit{l,~b})~(235$^{\circ}$,~~-9$^{\circ}$).

The depth of field was set to between 0.8333 and 2.0~mas (500 to 1200~pc), well beyond the expected position of CMa~OB2, to capture the full range of sources discussed in the literature. In order to only select sources with `good' astrometric solutions, i.e. eliminating very bright stars, multiple stars and those with extreme colours, \cite{lindegren2018re} recommends that the Re-normalised Unit Weight Error (RUWE) value associated with each source be taken in to account. For this study, only those sources with RUWE~$\leq$~1.4 are kept.

Additionally, all stars identified in our query results have positive parallaxes and values of $\sigma_{\varpi}$/${\varpi}$~<~0.08. We therefore define our distance values simply as an inverse parallax \citep[][p.17]{bailer2021estimating}. Also, only those sources with proper motions <10~mas~yr$^{-1}$ are retained (to exclude sources with excessively high proper motions).

The query identified 211652 sources.

\section{Discussion and Results} 
\label{Discussion}

\subsection{Identification of the OB population} 
\label{FindOB}

Previous studies of OB stars and their associations typically use the UBV photometric system \citep[e.g.][]{johnson1953fundamental}, although \cite{quintana2021revisiting} have recently used a multi-band approach in their study of the Cygnus OB associations. Few studies have been conducted identifying OB stars using \textit{Gaia} $G,~G_{bp}$ and $G_{rp}$ band photometry.

\cite{zari2021mapping} provide a method of identifying OBA stars using a cross-match query between \textit{Gaia} EDR3 and 2MASS photometry which involves an intermediate cross-match to \textit{Gaia} DR2 for sources with \textit{G} < 16~mag. This process requires photometric cuts in the data using J, G and $K_{s}$ colours.

\cite{pecaut2013intrinsic} present an analysis of the temperatures and intrinsic colours of pre-main sequence stars and a comparative table of magnitudes across a number of photometric bands for stars ranging from F0V to M9V \citep[][table~5]{pecaut2013intrinsic}. This table has subsequently been updated \citep{mamajek2019modern} to include \textit{Gaia} magnitudes and spectral classifications from O3V to Y4V. Within this table, OB stars are defined within the parameters $G_{(bp-rp)}$~<~0.05 and $mg$~<~1.0. Applying this filter to our data finds 1121 sources ($\sim$0.5\% of our data) within a well defined 'box' constrained by the filter limits.

Prior to the release of \textit{Gaia} DR3, filters such as these would be needed to analyse stellar data sets. With the release of DR3, spectral types are included in the \textit{Gaia} dataset and identified by the \textit{'spectraltype$\_$esphs'} flag in the \textit{'gaiadr3.astrophysical$\_$parameters'} folder. For our study, we choose to identify O and B-type stars solely through the use of \textit{Gaia} DR3 spectral types.

\begin{figure} 
    \includegraphics[width=\columnwidth]{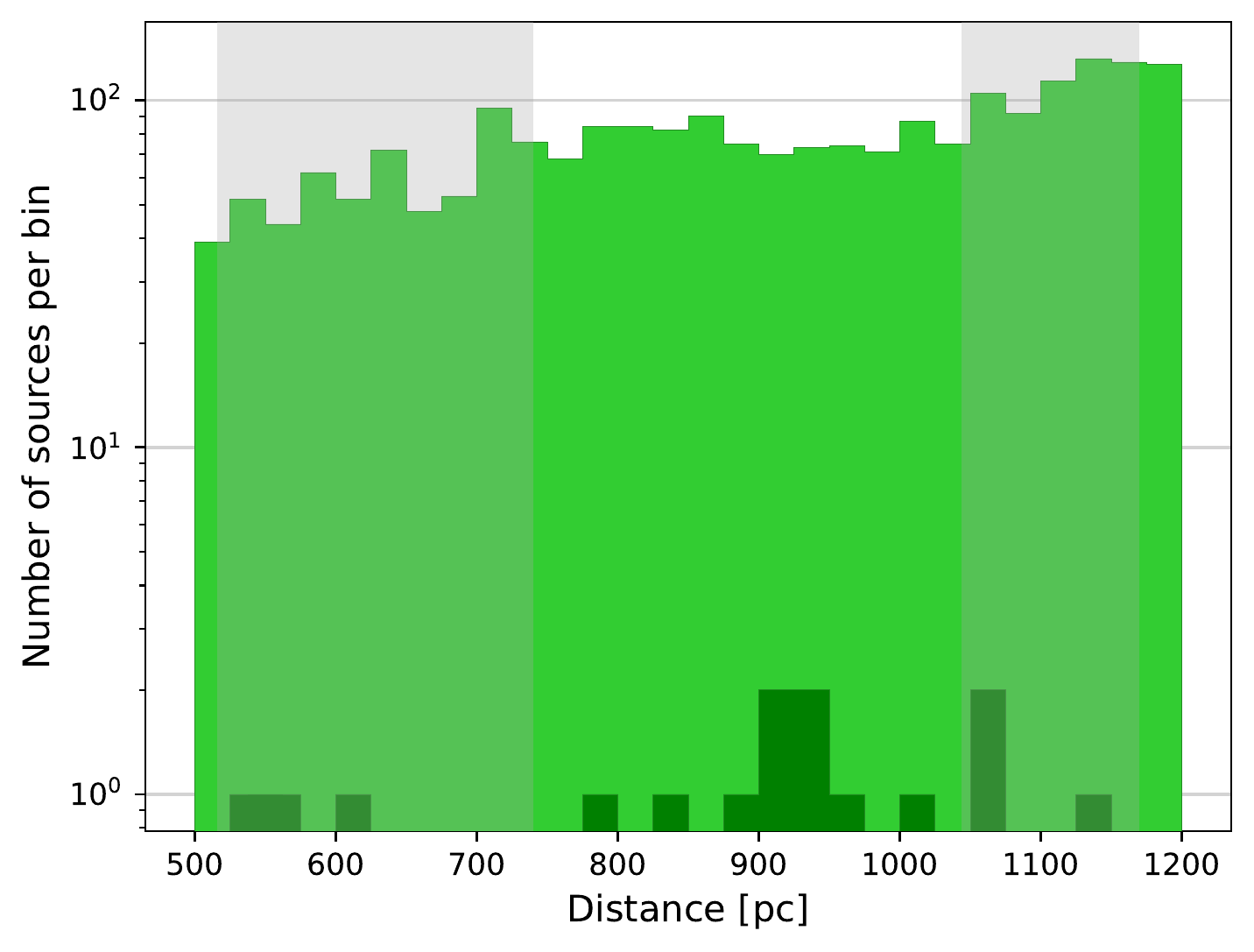}
    \caption[Histogram of \textit{Gaia} DR3 sources.]{The distance distribution of 2240 OB stars identified in our data using \textit{Gaia} DR3 spectral classifications. The vertical gray bands identify the bounds of the two associations described in the literature and detailed in Table~\ref{tab:Collinderstudies}, with the data binned at 25~pc. Note the use of a logarithmic y-axis to accentuate the distribution of O-type sources.}
    \label{fig:Coll121OBDist}
\end{figure}

Within our data set of 211652 sources, we find 15~O-type and 2225~B-type stars giving a total of 2240~stars, representing $\sim$1\% of our data, the distance distribution of which can be seen in Figure \ref{fig:Coll121OBDist}. It should be noted that the y-axis displays a logarithmic scale to accentuate the distribution of the O-type stars.

Assuming an almost near-constant density of the interstellar
medium on such a small scale, it is expected that the number of stars should
increase as R$^{2}$ mirroring the increase in cross-sectional area of each equal-
width distance bin. The increase in sources along the distance scale seen in our data is in agreement with this assumption.

Figure \ref{fig:Coll121OBDist} identifies the outer limits of the regions identified in the literature (see Table~\ref{tab:Collinderstudies}). At this scale it is not apparent that any coherent clusters are contained in the distribution of identified OB stars.

\begin{figure} 
    \includegraphics[width=\columnwidth]{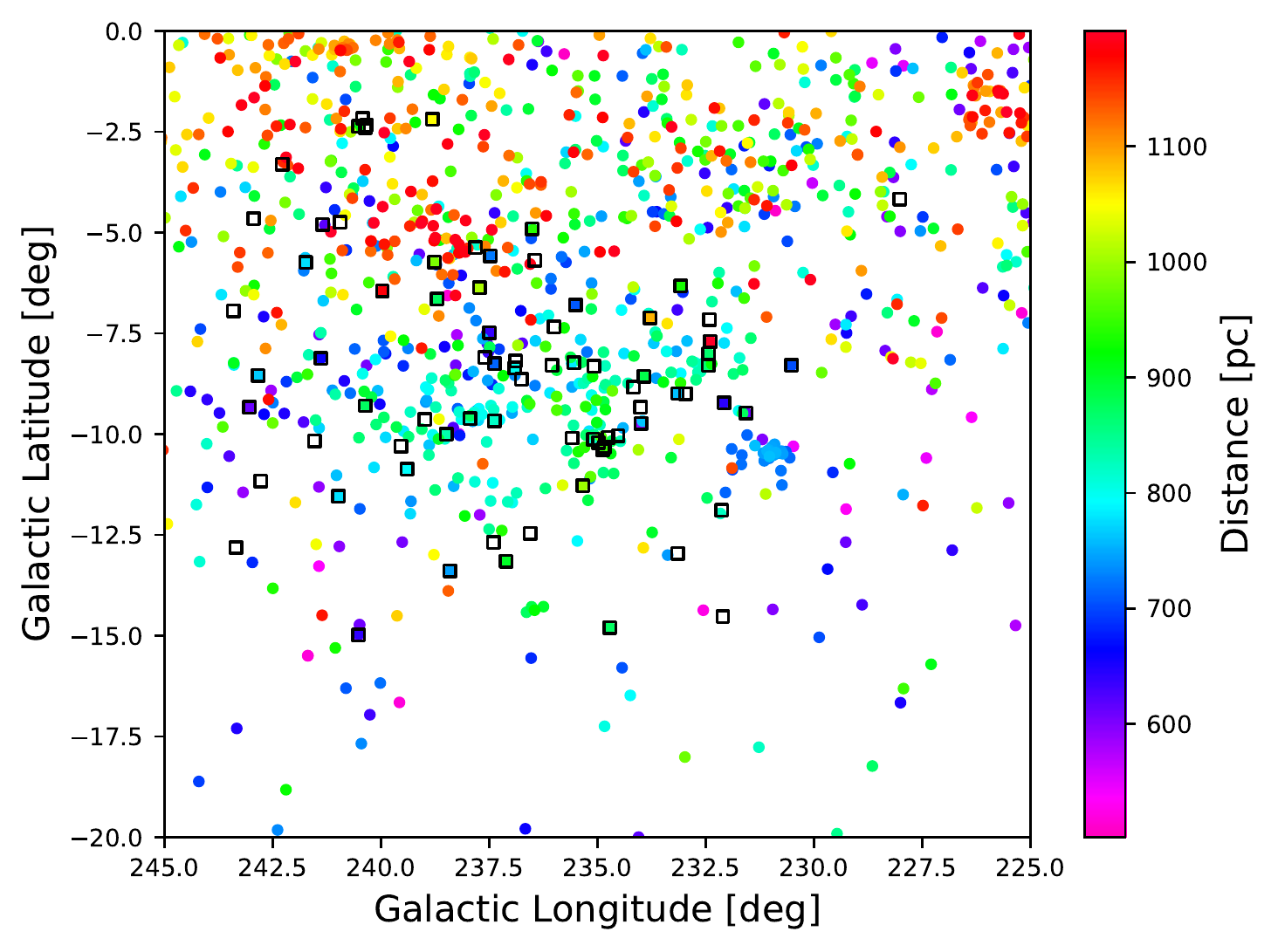}
    \caption[Combined sky plot of \textit{DR3} and literature sources.]{Combined positional sky plot showing both \textit{Gaia} DR3 and identified literature sources. \textit{Gaia} sources are shown colour coded against distance. The 78 literature sources from \cite{de1999hipparcos} and \cite{kaltcheva2000region} are shown as black open squares, of these, 26 are not coincident with our data.}
    \label{fig:CombinedSources}
\end{figure}

Figure~\ref{fig:CombinedSources} overlays the 78 sources identified in the \cite{de1999hipparcos} and \cite{kaltcheva2000region} studies, that are within our distance limits (shown as black open squares), onto a distance distribution of those sources identified in this study (Figure~\ref{fig:Coll121OBDist}). Of the 78 sources, SIMBAD\footnote{\href{http://simbad.u-strasbg.fr/simbad/}{http://simbad.u-strasbg.fr/simbad/}} identifies 8 as being spectral types other than OB.

It can be seen that sources identified in the previous studies are mainly clustered around $\sim$900~pc (green markers). The large grouping apparent at (\textit{l,~b})~(225$^{\circ}$,~~-1.5$^{\circ}$) lying at $\approx$1150~pc identifies the edge of the CMa~OB1 Association \citep[see e.g.][]{santos2021canis}. The further grouping at (\textit{l,~b})~(238$^{\circ}$,~~-5.5$^{\circ}$) may indicate the location of OB sources within NGC~2362 lying at $\approx$1300~pc \cite[e.g.][]{zhang2023distances}. The obvious over-density seen at (\textit{l,~b})~(231.0$^{\circ}$,~~-10.5$^{\circ}$), seen as a light blue distribution at approximately 750~pc, is the OB population of the M41 open cluster. This cluster, although in-field, does not form part of the Collinder~121 association. It is also apparent in this distribution, that a number of the literature OB sources do not fully align with sources identified in our study, as shown by the white square markers.

\begin{figure} 
    \includegraphics[width=\columnwidth]{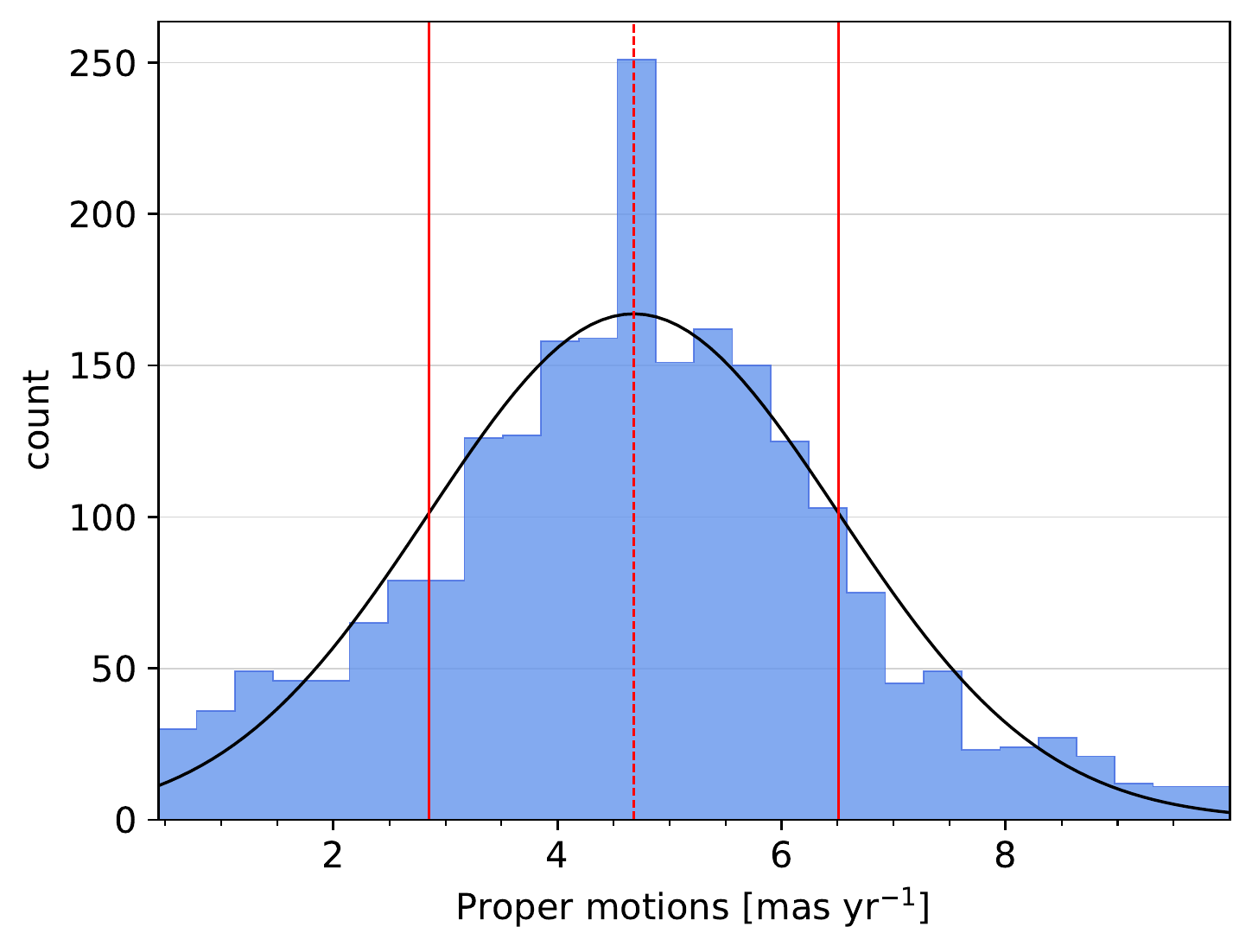}
    \caption[OB stars distribution of proper motions.]{The galactic proper motion distribution for the 2240 OB stars identified in this study. We find a mean value of proper motion ($\mu$) = 4.68$\pm$1.83~mas~yr$^{-1}$ to 1$\sigma$. Mean and 1$\sigma$ limits are shown by vertical red lines.}
    \label{fig:PMvectorHists}
\end{figure}

Having identified a significant number of OB sources, we now look at their proper motion distribution. Figure \ref{fig:PMvectorHists} shows a histogram of the proper motion for our OB population, the values from which are presented in Table~\ref{tab:PMStats}. For our study we only wish to consider those sources with the least variance from the mean, we therefore identify 866 sources within 1 standard deviation.

\begin{table} 
\centering
    \caption{Proper motion values for the OB stars shown in Figure~\ref{fig:PMvectorHists}. Mean and standard deviation values are given to 1 standard deviation (see text).}
    \label{tab:PMStats}
    \begin{threeparttable}
    \begin{tabular}{lccc}
    \hline
    \normalfont{Proper Motion} &
    \normalfont{Mean value [mas~yr$^{-1}$]} &
    \normalfont{SD [1$\sigma$]} \\
    \hline
{$\mu_\textit{l}*$} [mas~yr$^{-1}$] & -3.46 & 2.57 \\
{$\mu_\textit{b}$} [mas~yr$^{-1}$] & -1.93 & 1.70 \\
$\mu_{~total}$ [mas~yr$^{-1}$] & 4.68 & 1.83 \\
    \hline
    \end{tabular}
    \end{threeparttable}
\end{table}

For our population of 2240 stars, we find galactic proper motion mean values of {$\mu$} = 4.68$\pm$1.83~mas~yr$^{-1}$ with individual vectors of {$\mu_\textit{l}*$} = -3.46$\pm$2.57~mas~yr$^{-1}$ and {$\mu_\textit{b}$} = -1.93$\pm$1.70~mas~yr$^{-1}$ to 1$\sigma$. However, it must be remembered that this is a widespread population covering a distance range of 500~<~d~<~1200 pc and additional work is required in order to positively identify the existence of individual groups within the data.

\subsection{Clustering} 
\label{OBevidence}

To identify individual OB sub-groups within the population described, we perform 6-dimensional clustering analysis on \textit{Gaia}~DR3 values of \textit{l}, \textit{b}, distance (parallax derived), {$\mu_\textit{l}*$}, {$\mu_\textit{b}$} and radial velocity.

We chose to use the HDBSCAN clustering algorithm \citep{campello2013density, campello2015hierarchical}. It is part of a growing number of clustering algorithms and has been widely used within wide survey contexts  \citep[e.g.][]{kounkel2019untangling, logan2020unsupervised, kuhn2021spicy}.

HDBSCAN is a useful unsupervised learning tool in that it can identify groups with varying densities and shapes without the requirement to pre-specify the number of clusters expected in the data, as opposed to more commonly used algorithms such as K-Means \citep{macqueen1967classification} and spectral clustering \citep{ng2001spectral}. HDBSCAN can work with multi-dimensional data and builds a hierarchy of potential clusters, presenting the output as a spatial representation of the clusters as well as hierarchical tree diagrams.

There are two major, user defined parameters to consider when initialising HDBSCAN. These are the \textit{minimum samples} parameter which sets the minimum number of samples (data points) in the neighbourhood for a point to be considered a core point, effectively defining the density of the region, and the \textit{minimum cluster size} parameter defining the minimum number of points to be classified as a cluster. Determining these parameters is largely dependant on the output required and is data-dependant.

To determine optimal (although subjective) values of the above parameters, several runs of the HDBSCAN algorithm were conducted using values of \textit{minimum samples} between 8~-~30 and \textit{minimum cluster size} between 5~-~60. For this initial {\fontfamily{pcr}\selectfont Python} implementation of the HDBSCAN algorithm we use \textit{minimum samples}~=~15 to set the minimum number of sources required to form a cluster and \textit{minimum cluster size}~=~35 to minimise the presence of fake clusters.

\subsection{Searching for Moving Groups} 
\label{OBmoving}

\begin{figure} 
    \includegraphics[width=\columnwidth]{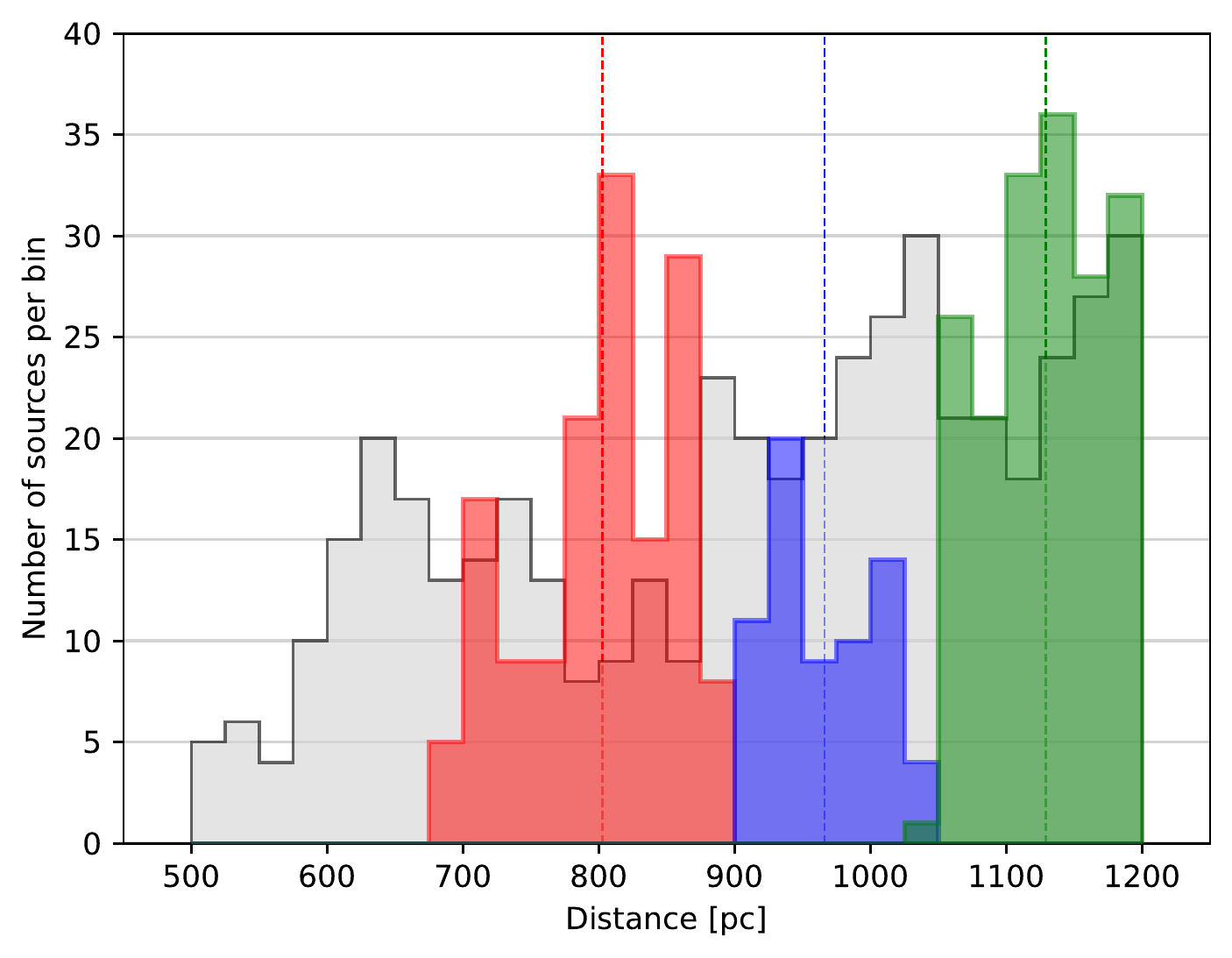}
    \caption[Distance distribution of 3 main groups.]{Distance distribution of 866 stars, binned at 25~pc, found using HDBSCAN. Group 1 (red) contains 146 members, centred on 803~$\pm$55~pc to 1$\sigma$. Group 2 (blue) lies at 966~$\pm$37~pc and contains 68 members. The third group (green) contains 177 members at a distance of 1129~$\pm$43~pc (see text). The vertical dotted lines identify the mean distance of each group. The grey area under the black line identifies the distribution of those sources which did not meet the group membership threshold (see text).}
    \label{fig:3groupDist}
\end{figure}

HDBSCAN is run on the \textit{Gaia} catalogue previously described, namely $\pm$1$\sigma$ values of {$\mu_\textit{l}*$} and {$\mu_\textit{b}$} derived from Table~\ref{tab:PMStats} and distance values~>500~pc and <1200~pc, we therefore use our previously identified 866 stars (see Section~\ref{FindOB}). Three groups of stars are found by an initial run and their positions are shown in Figure~\ref{fig:3groupDist}. We refer to these groups as Group 1, 2, and 3 in order of increasing distance and respectively colour their stars red, blue and green in subsequent plots.

The nearest group, lying at a mean distance of 803~$\pm$55~pc to 1$\sigma$ contains 146 members centred on (\textit{l,~b})(236.4$^{\circ}$,~-6.52$^{\circ}$). The two other groups, centred on (\textit{l,~b})(235.13$^{\circ}$,~-2.74$^{\circ}$) and (\textit{l,~b})(236.3$^{\circ}$,~-0.73$^{\circ}$) contain 68 and 177 members respectively and are located at 1$\sigma$ mean distances of 966~$\pm$37~pc and 1129~$\pm$43~pc respectively (see Figure~\ref{fig:3groupDist}). The remaining 475 sources do not pass the significance threshold to be included in the identified groups and are shown by the grey region in Figure~\ref{fig:3groupDist}.

The peculiar velocity of the Sun with respect to the Local Standard of Rest (LSR) (U{$_\odot$}, V{$_\odot$},W{$_\odot$})$_{LSR}$ plays a significant role in the analysis of  Galactic stellar kinematics. To properly analyze Galactic orbits, solar peculiar velocity should be removed from the observed velocities of stars, since it only describes the Solar orbit.

To remove the rotation effects due to the peculiar velocity of the Sun with respect to the LSR, we used the values derived in \cite{bobylev2015new}, namely (\textit{U}{$_\odot$}, \textit{V}{$_\odot$}, \textit{W}{$_\odot$})$_{LSR}$~=~(6.0, 10.6, 6.5)~km s$^{-1}$ determined through a study of young stars in the Solar neighbourhood using \textit{Hipparcos} data. This study was chosen since it was based on a similar population of stars, at a comparable distance to our own study.

\begin{figure} 
    \includegraphics[width=\columnwidth]{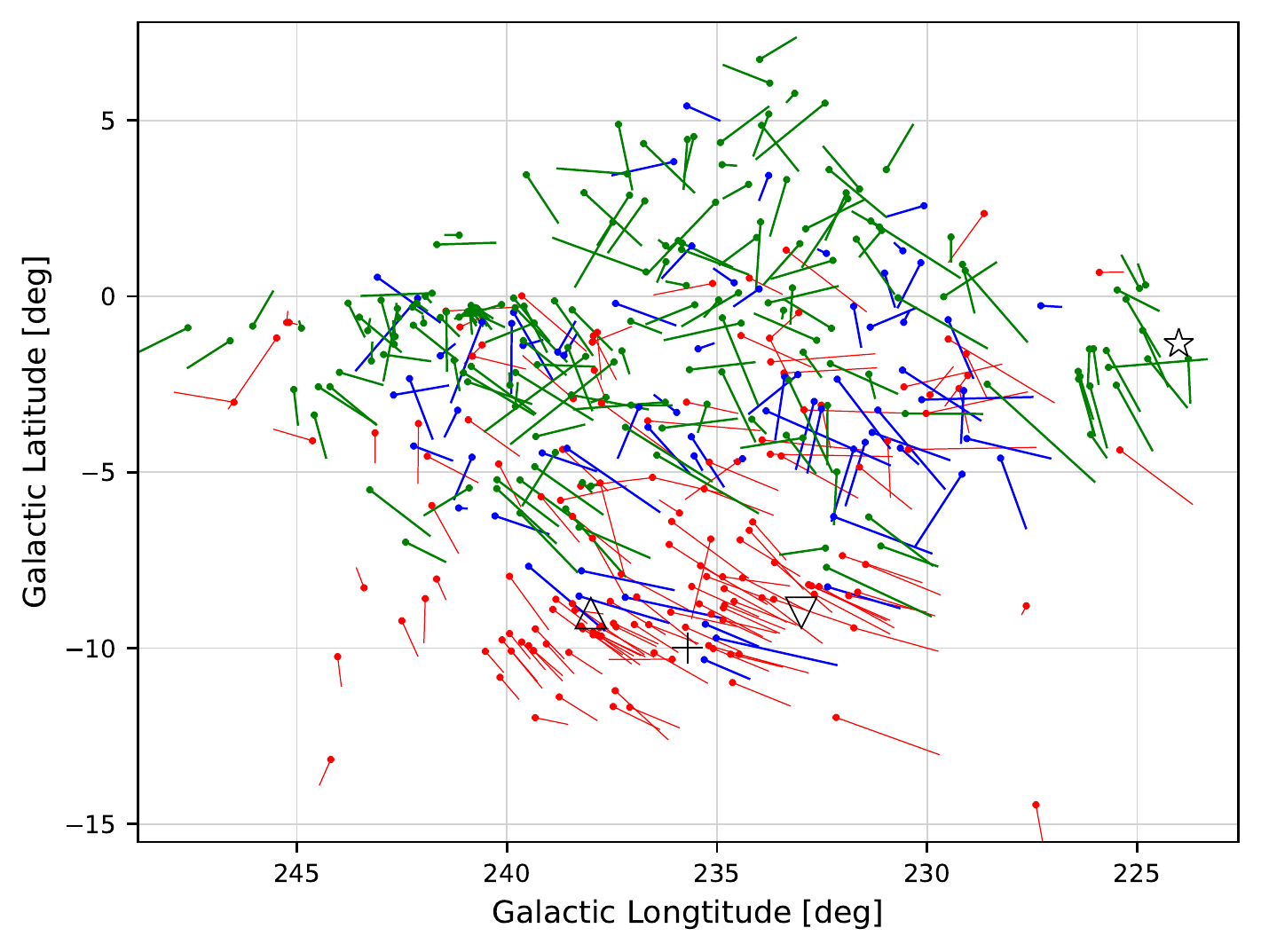}
    \caption{Corrected velocity vectors for the three main groups identified. The locations of the main clusters identified in the literature are shown for interest, as is the central location of the Collinder~121 region, denoted by the black cross. The coherent vectors of group 1 lying in the plane around \textit{b}~=-10$^{\circ}$ are easily identifiable. The colour coding is consistent with Figure~\ref{fig:3groupDist}.}
    \label{fig:3groupVectors}
\end{figure}

Plotting the corrected proper motion velocity vectors of our three groups (Figure~\ref{fig:3groupVectors}) shows little coherence in the movement of the two farthest groups (Groups 2 and 3), whilst there appears to be a uniformity of motion in the region surrounding the recognised centre of Collinder~121 belonging to Group 1. Identified clusters from the literature are shown for comparison, CMa~OB1 is denoted by a star, CMa~OB2 by a downward facing triangle and Cr~121 by an upward pointing triangle. Closer inspection of the vectors in Group 1 suggests that there are small differences in the direction of motion along the longitudinal scale.

\begin{figure} 
    \includegraphics[width=\columnwidth]{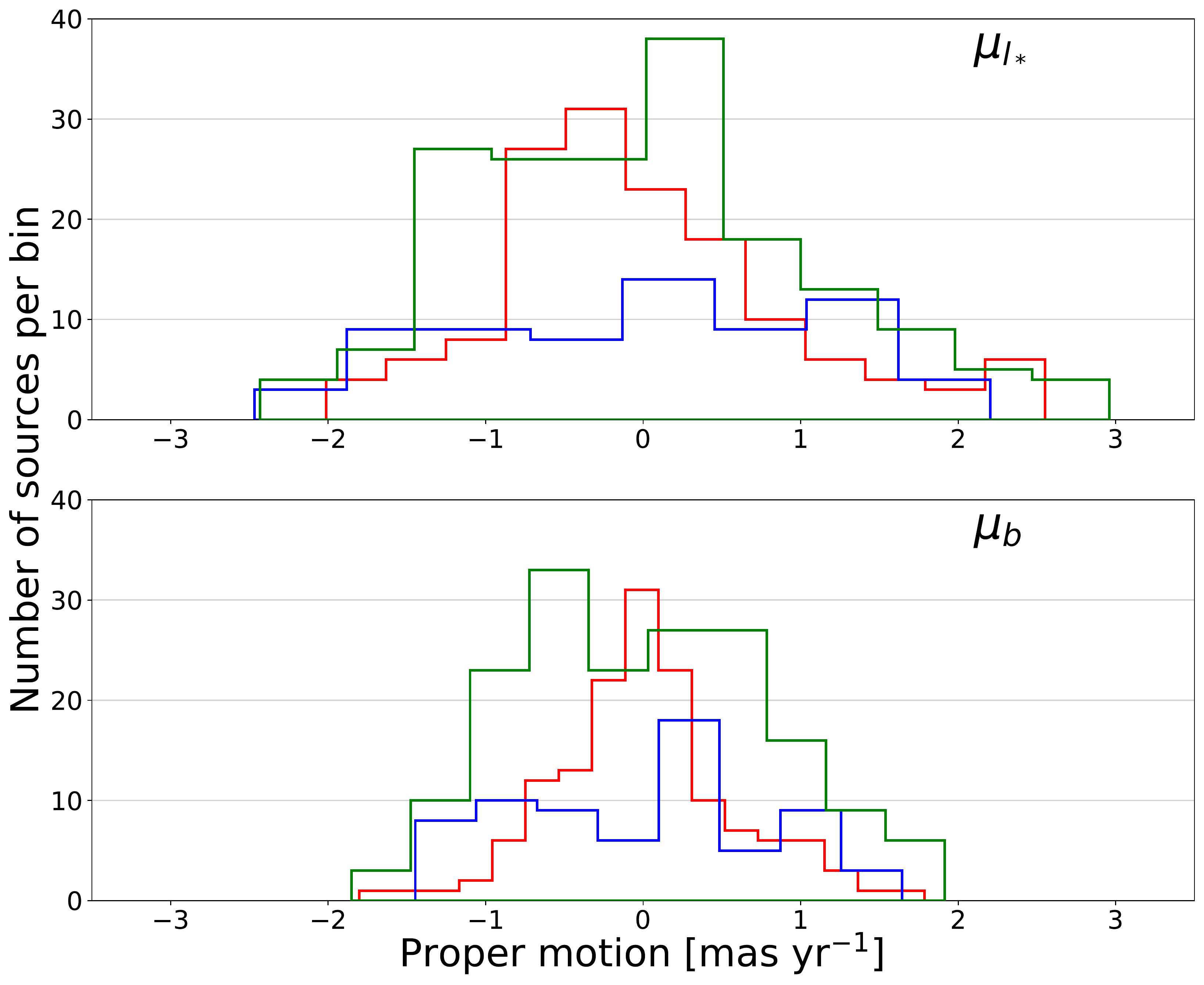}
    \caption{Distribution of mean subtracted, corrected {$\mu_\textit{l*}$} (top figure) and {$\mu_\textit{b}$} (bottom figure) proper motions after removal of the regional bulk motion due to solar motion. The group colour codes are consistent with those used in Figure~\ref{fig:3groupDist}.}
    \label{fig:3groupProperMotions}
\end{figure}

To obtain a clearer understanding of the motions within each group we found their mean motions and subtracted this value from the motion of individual group members. This resultant mean subtracted motion (Figure~\ref{fig:3groupProperMotions}) provides a direct comparison between the motions of each group and removes the apparent randomness of motion depicted in (Figure~\ref{fig:3groupVectors}).

\subsection{Identifying substructure} 
\label{subgroupd}

To further investigate the possible substructure of the 146 members within Group 1 we again choose only those sources with the least variance from the mean and remove outlying sources by reducing our distance parameter to the 1$\sigma$ values determined above. We reduce our spatial scales to concentrate on the over-densities in the central region, this reduces the number of sources to 77. It is recognised that a less stringent distance cut may have significantly increased this number. With this reduced source list we conducted a second HDBSCAN search, this time performing a 4-dimensional analysis using values of  \textit{l} and \textit{b} and previously determined values of {$\mu_\textit{l}*$} and {$\mu_\textit{b}$} corrected for solar motion.

\begin{figure} 
    \includegraphics[width=\columnwidth]{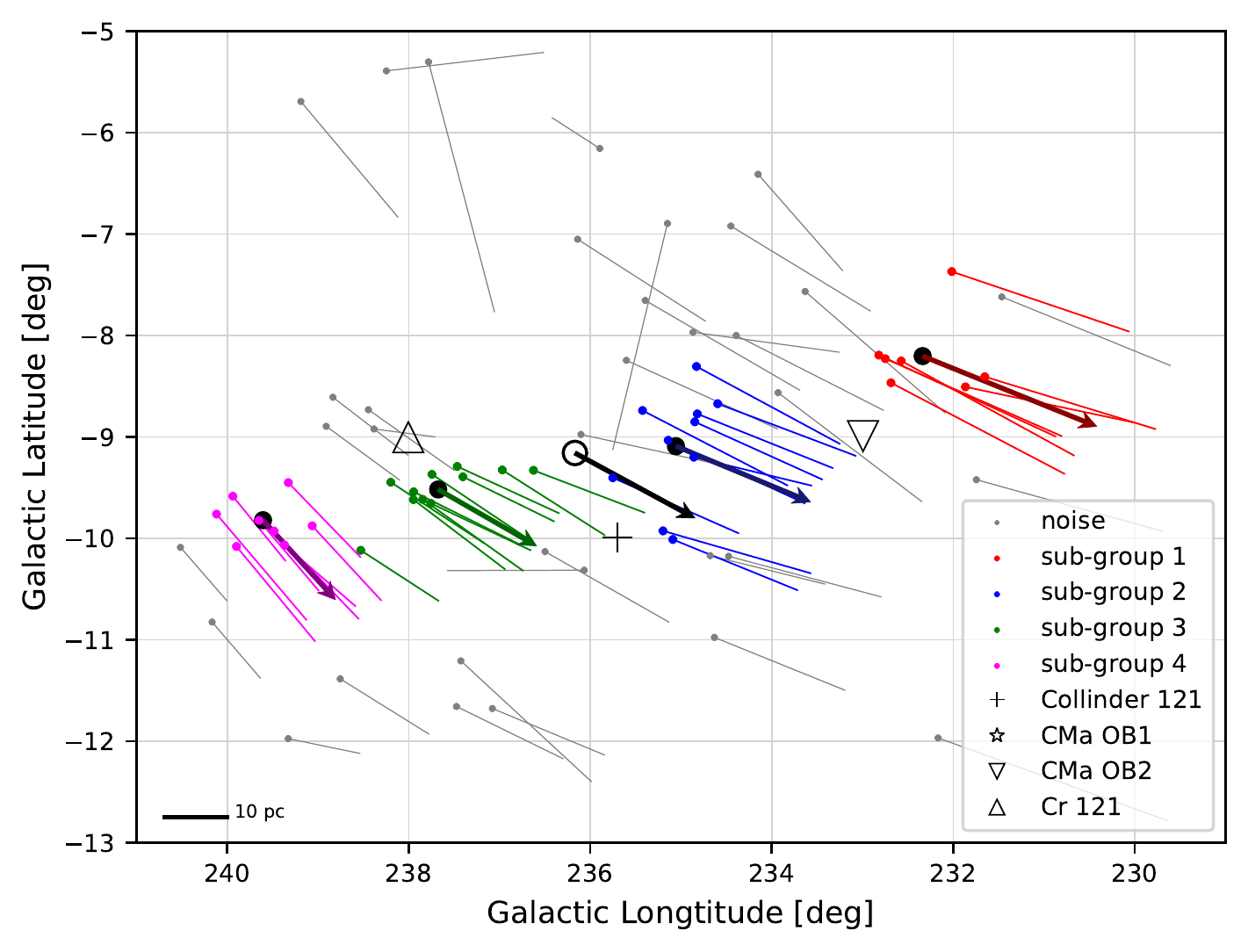}
    \caption[Subgroup mean proper motion vectors.]{Results of performing clustering analysis of the group 1 members identified in Figure~\ref{fig:3groupVectors} (see text). Four distinct subgroups are identified with similar directions of motion. The location and overall mean velocity of all subgroups is shown as a black open circle vector at (\textit{l,~b})(~236.17$^{\circ}$,~-9.16$^{\circ}$). The grey vectors are those sources not allocated to one of the subgroups by the clustering algorithm. It is evident that many of these vectors are in-line with those of the subgroups and may be indicative of a larger association. A noticeable (figure) left to right increase in the size of the vectors exists in the longitudinal axis. The distance scale at bottom left is set for the mean distance of the subgroups at 827~pc. The parameters of each subgroup are contained in Table~\ref{tab:SubGroups}.}
    \label{fig:SubGroupVectors}
\end{figure}

Reducing our clustering algorithm parameters to \textit{min\_samples}~=~5 and \textit{min\_cluster\_size}~=~4, to identify small-scale independent associations, and applying these parameters to a further HDBSCAN algorithm finds 36 sources in four subgroups (Figure~\ref{fig:SubGroupVectors}), the parameters of which are given in Table~\ref{tab:SubGroups}. Of the 77 sources, 41 are considered to lie outside the identified subgroups and are shown, with their vectors, as grey sources in Figure~\ref{fig:SubGroupVectors}.

\begin{figure} 
    \includegraphics[width=\columnwidth]{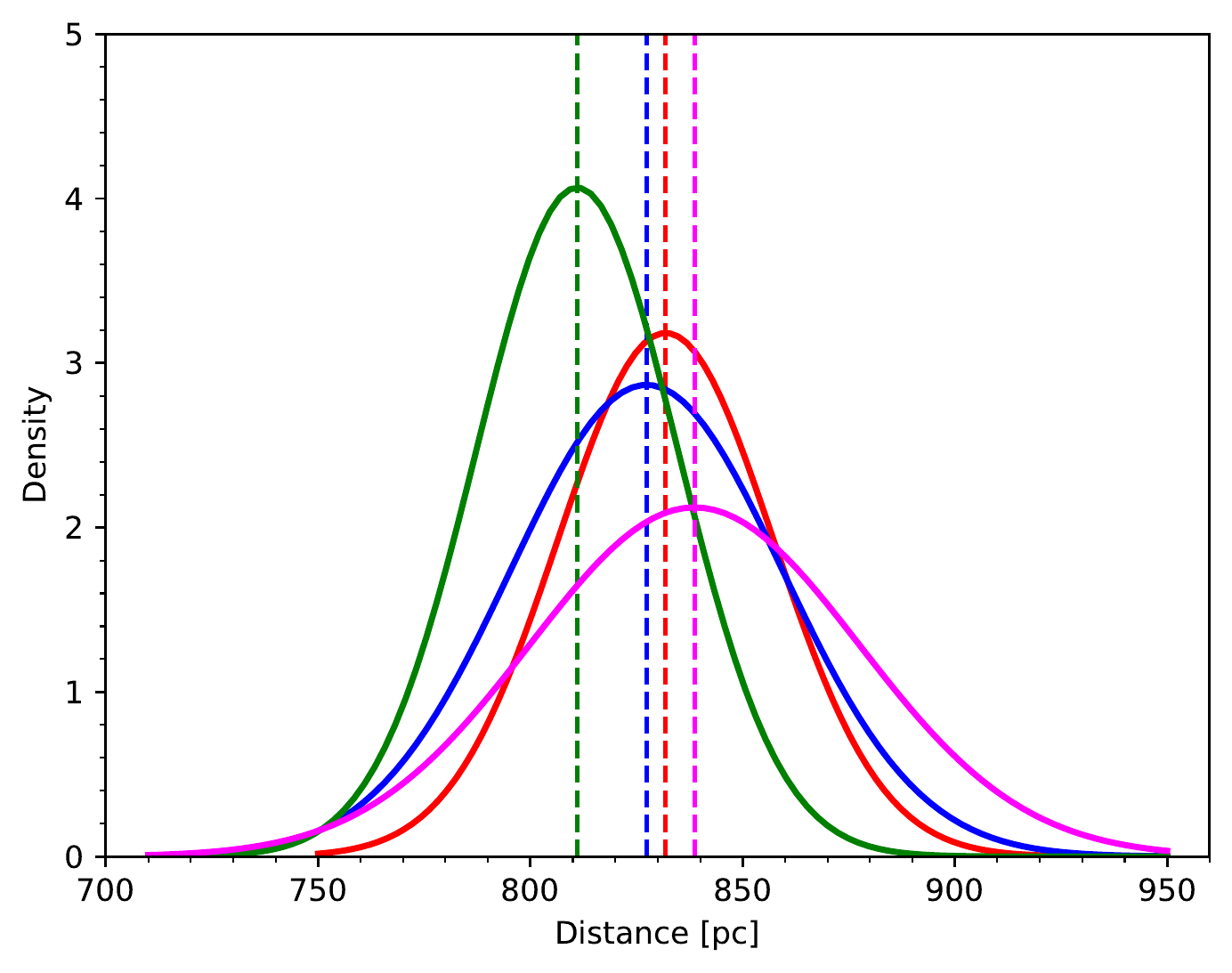}
    \caption[Subgroup distance distributions.]{Location of each subgroup in distance space. Colour coding of the groups is consistent with Figure~\ref{fig:SubGroupVectors}.}
    \label{fig:SubGroupDist}
\end{figure}

\begin{table} 
    \caption{Solar corrected proper motion vector values for the OB star subgroups shown in Figure~\ref{fig:SubGroupVectors}.}
    \label{tab:SubGroups}
    \begin{threeparttable}
    \begin{tabular}{lccc}
    \hline
    \normalfont{Subgroup} & {Number} & {$\mu_{~mean}$ Corrected} & {Angle (E of N)}\\
    {} & {in subgroup} & [mas~yr$^{-1}$] & [degrees]\\
    \hline
1 & 7 & 2.04 & 248.00 \\
2 & 10 & 1.58 & 247.50 \\
3 & 11 & 1.22 & 240.16 \\
4 & 8 & 1.11 & 222.35 \\

\hline
Mean & & \\
{Group Motion} & & 1.46 & 239.50 \\
    \hline
    \end{tabular}
    \end{threeparttable}
\end{table}

\begin{figure} 
    \includegraphics[width=\columnwidth]{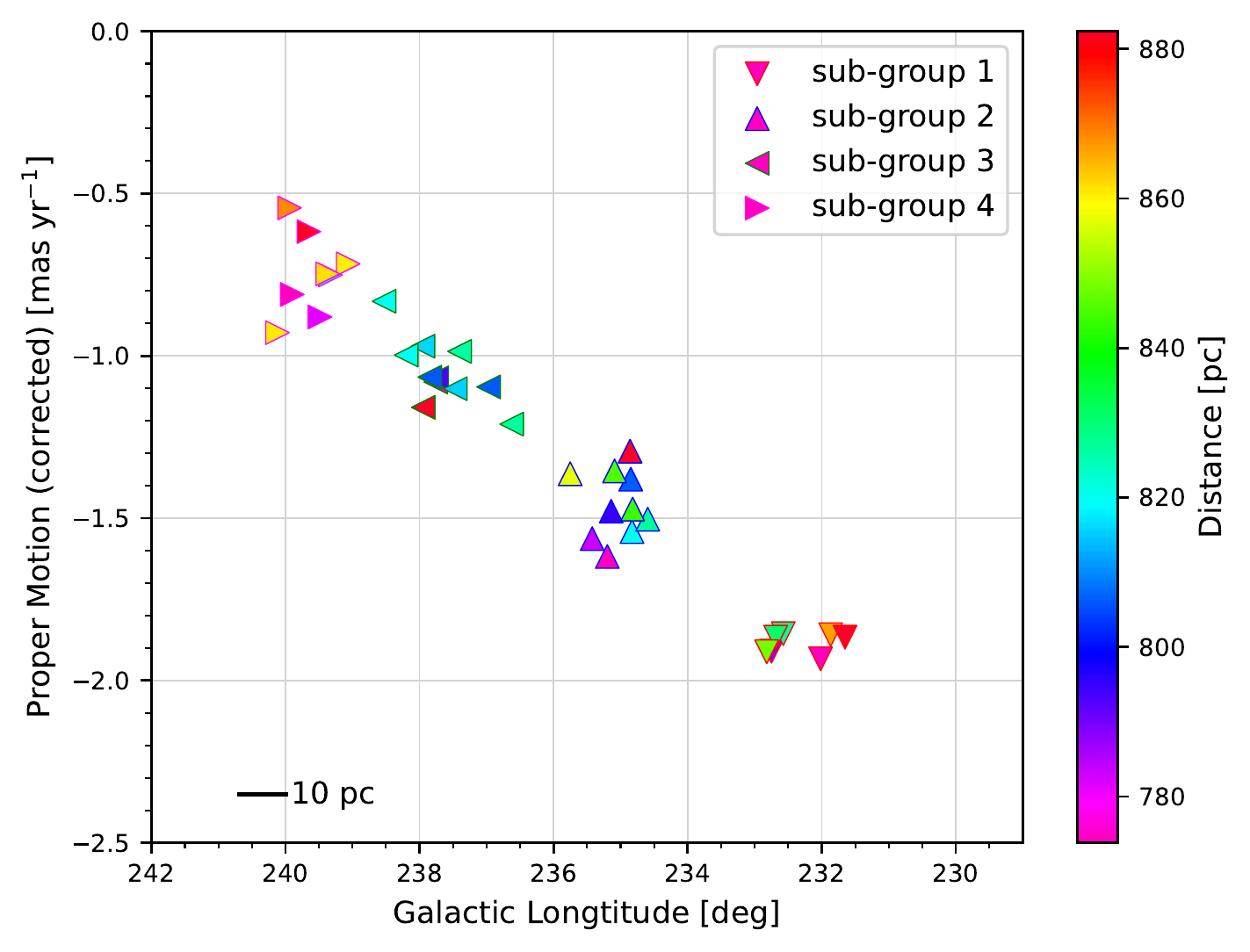}
    \caption[Kinematic pattern of subgroups.]{Galactic longitude as a function of corrected proper motions in galactic longitude for the identified subgroups showing the large-scale kinematic pattern. The extent to which each subgroup is separated along the line-of-sight is highlighted by their colour distributions. The distance scale at bottom left is set for the mean distance of the subgroups at 827~pc.}
    \label{fig:SubGroupGradient}
\end{figure}

Interestingly, Figure~\ref{fig:SubGroupVectors} shows a noticeable increase in the size of the corrected velocity vectors moving from left to right along the longitudinal axis. This large-scale kinematic pattern between proper motion and position (see Figure~\ref{fig:SubGroupGradient}) has also been observed in other studies of OB stars in Cygnus \citep{quintana2021revisiting, quintana2022large} and in the study of OB star proper motions in the Carina Arm \citep{drew2021proper}. The precise cause of this correlation between \textit{l} and \textit{$\mu_{l}$} is unclear but previous studies have suggested that it may be caused by feedback processes from a previous generation of stars. In this study we have corrected for solar motion with respect to the LSR, we therefore suggest that this proper motion pattern may simply be due to systematic differential galactic rotation as indicated by the calculations of \cite{vaher2020recipe}.

Figure~\ref{fig:SubGroupDist} provides individual distance distributions for the four subgroups identified, the parameters of which are given in Table~\ref{tab:DistStats}. For these subgroups, their mean distance lies at 827$\pm$30~pc.

\begin{table} 
\centering
    \caption{Distance parameters of the 4 subgroups identified in the central Collinder~121 region shown in Figure~\ref{fig:SubGroupDist}. Mean and 1$\sigma$ standard deviation values are provided. Details of individual group members can be found in Appendix~\ref{DR3 4subgroups}, Table~\ref{tab:chartable}.}
    \label{tab:DistStats}
    \begin{threeparttable}
    \begin{tabular}{lccc}
    \hline
    \normalfont{Subgroup} &
    \normalfont{Distance [pc]} &
    \normalfont{SD [1$\sigma$]} \\
    \hline
1 & 831.87 & 25.34 \\
2 & 827.37 & 32.05 \\
3 & 811.03 & 23.78 \\
4 & 838.66 & 38.82 \\

\hline
Mean & 827.23 & \\
    \hline
    \end{tabular}
    \end{threeparttable}
\end{table}

Whilst individual group numbers are small it can be seen that they are all coincident in their line of sight distances even though their velocities and directions of travel vary. From these indications we suggest that our subgroups may in fact be members of a single larger association that HDBSCAN has failed to characterise, members of which may be beyond the spatial distribution of the sample studied.

\subsection{Isochronal Ages} 
\label{OBAges}

In their study of stellar evolutionary tracks and isochrones, \cite{bressan2012parsec} indicate that the ages of OB association members should be in the range 5–10~Myr with stars in class O8 being younger than 5~Myr and those in class B0 younger than 10~Myr. In this context it is interesting to identify the isochronal ages of the sub-group members identified in this study.

There are a number of web-based packages available from which stellar evolutionary models may be obtained, notably the PARSEC\footnote{\href{http://stev.oapd.inaf.it/cgi-bin/cmd}{http://stev.oapd.inaf.it/cgi-bin/cmd}} and Geneva\footnote{\href{https://www.unige.ch/sciences/astro/evolution/en/database/syclist/}{https://www.unige.ch/sciences/astro/evolution/en/database/syclist/}} libraries. Both of these libraries have the option of selecting \textit{Gaia} photometric systems.

The Geneva ‘high mass-loss’ isochrones \citep{meynet1994grids} are
mainly chosen for the modelling of starbursts \citep[e.g.][]{levesque2010theoretical, byler2017nebular} since they accurately model the stages of low-luminosity Wolf–Rayet stars.

The PARSEC~v1.2S library provides a better treatment of boundary conditions in low-mass stars \citep[(M~$\leq$~0.6M{$_\odot$})][]{chen2014improving} and the mass-loss rates for more massive stars \citep[(14~$\leq$~M$/$M$\odot~\leq$~350)][]{chen2015parsec}.

\begin{figure} 
    \includegraphics[width=\columnwidth]{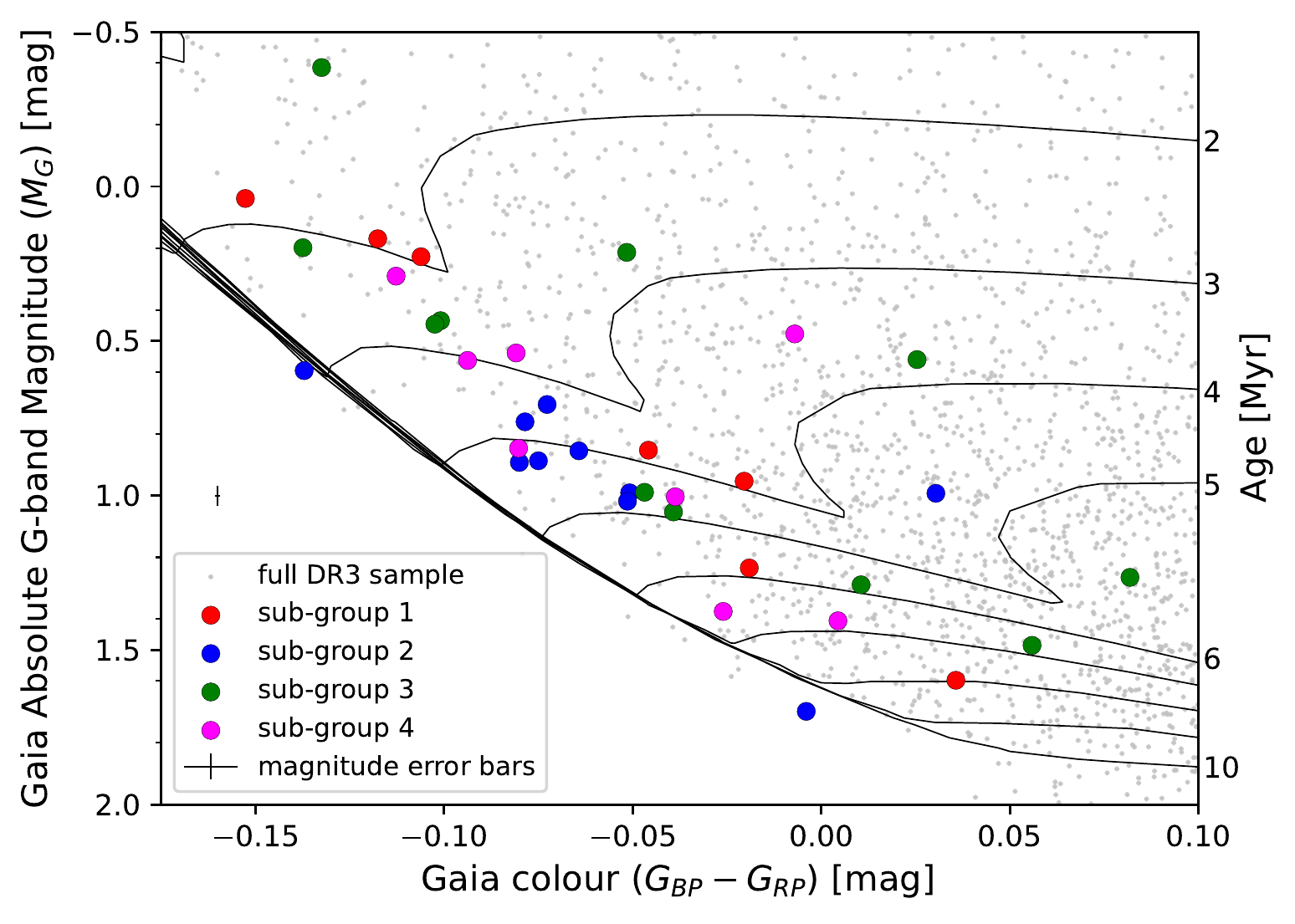}
    \caption[Subgroup CMD.]{Colour magnitude diagram showing the location of subgroup members, against a background of all query identified objects. PARSEC isochrones have been overlaid on the CMD at intervals ranging from 2.0~Myr to 10.0~Myr. Note also, the inclusion of magnitude error bars which are negligible in comparison to the source locations (see text).}
    \label{fig:SubGroupCMD}
\end{figure}

Due to its ability to generate isochrones for stars in our mass range, we use the PARSEC library with the following options: YBC+new~Vega photometric system; no circumstellar dust; total extinction A$_V$~=~0.2126 mag and a Kroupa initial mass function. Figure~\ref{fig:SubGroupCMD} places the members of our four sub-groups on PARSEC isochrones that are corrected for use with \textit{Gaia} photometry.

The extinction information provided in the DR3 archive is (naturally) provided in terms of A$_G$, whereas the Padova PARSEC input form requires interstellar extinction to be input in terms of A$_V$. The conversion between A$_G$ and A$_V$ is not a simple matter as discussed in \cite{andrae2018gaia} and \cite{evans2018gaia}. In their study, \cite{matsunaga2017time} show that G-band extinction values are comparatively smaller than those in the V-band and almost imperceptible for values <~1~mag.

The mean \textit{Gaia} DR3 value of extinction for the subgroup members as given by the {'ag\_gspphot'} field is 0.06675~mag and we use this as the corresponding value of A$_V$ for use in plotting the PARSEC isochrones.

Also, in presenting PARSEC isochrones it should be noted that the PARSEC models use the relation $A_{G}\sim$2~*~E($G_{BP}$~\textminus~$G_{RP}$) built into them, thereby accounting for the line-of-sight extinction and reddening of the \textit{Gaia} $G_{BP}$~\textminus~$G_{RP}$ colour \citep[see][]{andrae2018gaia}.

We see from Figure~\ref{fig:SubGroupCMD} that the majority of subgroup members have isochronal ages younger than $\sim$5~Myr and that there is no clear peer grouping within the age distributions. Two of the sources lie to the left of the Main Sequence suggesting that they appear bluer in the \textit{Gaia} passbands and may in fact contain circumstellar disks with the excess blue colour resulting from light scattering from a highly inclined disk \citep[see e.g.][]{robitaille2006interpreting}.

Magnitude mean error bars have been included in Figure~\ref{fig:SubGroupCMD} for both M$_G$ (3.40~x~10$^{-2}$~mag) and G$_{BP}$-G$_{RP}$ colour (7.00~x~10$^{-4}$~mag) for all subgroup members. The reason that only mean error bars are shown is that the errors across this set of stars is so low. In all instances, the size of the plotted marker symbols is comparable to the size of the error bars. More specifically, the M$_G$ and G$_{BP}$-G$_{RP}$ errors are: 4.12~x~10$^{-2}$~mag, 1.40~x~10$^{-3}$~mag; 3.30~x~10$^{-2}$~mag, 5.00~x~10$^{-4}$~mag; 3.22~x~10$^{-2}$~mag, 7.00~x~10$^{-4}$~mag and 3.15~x~10$^{-2}$~mag, 6.00~x~10$^{-4}$~mag in order for subgroups 1, 2, 3 and 4 respectively.

\section{Summary} 
\label{Summary}

We re-examine the literature surrounding Collinder~121 with the improved distance measurements from \textit{Gaia}. What we find is that  our understanding of the characteristics of the central region of Collinder~121 have changed dramatically with the advent of the \textit{Gaia} era.

Using \textit{Hipparcos} parallax derived distances, previous studies of this region have indicated the existence of an OB association at approximately 628~pc and a possible more distant open cluster in the order of 1109~pc (see Table~\ref{tab:Collinderstudies}).

We find no evidence for the open cluster at 1085~pc in the central region of Collinder~121 as identified by \cite{kaltcheva2000region}. We do however find a large divergent group (Group~3), located at approximately 1129~pc centered around (\textit{l,~b})(~236.3$^{\circ}$,~-0.73$^{\circ}$) which shows little coherence in movement (see Figure~\ref{fig:3groupVectors}). Considering the now well understood limitations in \textit{Hipparcos} parallax values at these distances, the identification of additional clustering in the \textit{Gaia} data supports the existence of other associations in the region, although not in the specific areas or at the distances reported in the literature.

The identification of a population of stars in this study, centred on (\textit{l,~b})(236.4$^{\circ}$,~-6.52$^{\circ}$) lying at a mean distance of 803~pc, surrounding the centre of Collinder~121, supports the general findings of previous studies but now provides a more certain measure of its location due to the microarcsecond measurements made by \textit{Gaia}.

Within this population we find four coherent sub-groups with a combined corrected mean proper motion of 1.46~mas~yr$^{-1}$ moving with a mean vector of 239.5~degrees East of North, centred on (\textit{l,~b})(~236.17$^{\circ}$,~-9.16$^{\circ}$), which may be indicative of a larger moving group.

\section*{Acknowledgements} 

We have used data from the European Space Agency (ESA) mission \textit{Gaia} (\href{https://www.cosmos.esa.int/gaia}{https://www.cosmos.esa.int/gaia}), processed by the \textit{Gaia} Data Processing and Analysis Consortium (DPAC) (\href{https://www.cosmos.esa.int/web/gaia/dpac/consortium}{https://www.cosmos.esa.int/web/gaia/dpac/consortium}). Funding for the DPAC has been provided by national institutions, in particular the institutions participating in the \textit{Gaia} Multilateral Agreement.

This work has used the NASA Astrophysics Data System (ADS) Bibliographic Services (\href{http://ads.harvard.edu/}{http://ads.harvard.edu/}) as well as the VizieR catalogue access tool (\href{http://vizier.u-strasbg.fr/viz-bin/VizieR}{http://vizier.u-strasbg.fr/viz-bin/VizieR}) and SIMBAD astronomical database (\href{http://simbad.u-strasbg.fr/simbad/}{http://simbad.u-strasbg.fr/simbad/}), operated at CDS, Strasbourg, France. Figure 1 has been sourced from the Aladin Sky Atlas at \href{https://aladin.u-strasbg.fr/}{https://aladin.u-strasbg.fr/} developed and maintained by the Centre de Données astronomiques de Strasbourg.

This research has made use of {\fontfamily{pcr}\selectfont Astropy}, a community-developed core {\fontfamily{pcr}\selectfont Python} (\href{https://www.python.org/}{https://www.python.org/}) module for Astronomy \citep{robitaille2013astropy, price2018astropy}.  This work has also made extensive use of {\fontfamily{pcr}\selectfont Matplotlib} \citep{hunter2007matplotlib}, {\fontfamily{pcr}\selectfont SciPy} \citep{van2014scikit} and {\fontfamily{pcr}\selectfont NumPy} \citep{van2011numpy}. This work would not have been possible without the countless hours put in by members of the open-source community around the world.

The colour magnitude diagram has been produced using isochrones generated by version 3.6 of the PARSEC web interface, last modified on 19-Jul-2022. This service is maintained by Léo Girardi at the Osservatorio Astronomico di Padova.

We are grateful to the anonymous referee for their supportive and constructive comments that helped us improve the manuscript.

\section*{Data Availability} 

We have used data from the European Space Agency (ESA) mission \textit{Gaia} (\href{https://www.cosmos.esa.int/gaia}{https://www.cosmos.esa.int/gaia}), processed by the \textit{Gaia} Data Processing and Analysis Consortium (DPAC) (\href{https://www.cosmos.esa.int/web/gaia/dpac/consortium}{https://www.cosmos.esa.int/web/gaia/dpac/consortium}). The data underlying this article are available through the \textit{Gaia} Archive, at \href{https://gea.esac.esa.int/archive/}{https://gea.esac.esa.int/archive/}.


\bibliographystyle{mnras}
\bibliography{Frodo_Bibliography} 



\appendix 

\section{Subgroup Properties of \textit{G\lowercase{aia}} DR3 Collinder~121 sub-groups.}
\label{DR3 4subgroups}

Table~\ref{tab:chartable} contains details of the 4 subgroups identified in this study. Where proper motion values are given, these are the established values given in the \textit{Gaia} Archive .

\begin{table*}
    \caption{Subgroup Properties of \textit{Gaia} DR3 sources in the region of Collinder~121.}
    \label{tab:chartable}
    \begin{tabular}{lccccccccc}
    \hline
    \textit{Gaia} DR3 source$\_$id & RA & Dec & l & b & Parallax & Distance & {$\mu_\textit{l}*$} & {$\mu_\textit{b}$} & $\mu_{Total}$ \\
    {} & [deg] & [deg] & [deg] & [deg] & [mas] & [pc] & [mas yr$^{-1}$] & [mas yr$^{-1}$] & [mas yr$^{-1}$] \\
    \hline
Subgroup 1 ({Red}) & & & &  &  &  &  &  & \\
2932106969639732864 & 104.9191 & -20.3090 & 232.0169 & -7.3695 & 1.26 & 792.34 & -4.48 & -1.59 & 4.75 \\
2931905174892889344 & 104.4357 & -21.3381 & 232.7487 & -8.2282 & 1.25 & 802.12 & -4.58 & -1.67 & 4.87 \\
2931911084769537280 & 104.3332 & -21.1915 & 232.5733 & -8.2499 & 1.21 & 828.94 & -4.51 & -1.81 & 4.85 \\
2931901395321506944 & 104.1776 & -21.3867 & 232.6865 & -8.4657 & 1.20 & 831.62 & -4.54 & -1.74 & 4.86 \\
2931892251327971072 & 104.5033 & -21.3866 & 232.8206 & -8.1929 & 1.19 & 843.71 & -4.61 & -1.65 & 4.90 \\
2932187062184721152 & 103.7589 & -20.6743 & 231.8661 & -8.5062 & 1.17 & 856.74 & -4.41 & -1.12 & 4.55 \\
2932288767012868480 & 103.7568 & -20.4404 & 231.6525 & -8.4059 & 1.15 & 867.63 & -4.39 & -1.29 & 4.57 \\
\hline \hline \\
Subgroup 2 (Blue) & & & &  &  &  &  &  & \\
2922297573567890944 & 103.9142 & -24.2542 & 235.1986 & -9.9260 & 1.28 & 778.94 & -4.67 & -1.15 & 4.81 \\
2921654805945282816 & 105.1921 & -23.9413 & 235.4243 & -8.7389 & 1.27 & 788.03 & -4.66 & -1.61 & 4.93 \\
2922416110361955456 & 104.7698 & -23.8173 & 235.1411 & -9.0325 & 1.25 & 799.36 & -4.53 & -1.44 & 4.76 \\
2922444049133276288 & 104.8110 & -23.4778 & 234.8483 & -8.8508 & 1.23 & 810.76 & -4.39 & -1.38 & 4.60 \\
2922465489610732800 & 105.3368 & -23.2226 & 234.8296 & -8.3052 & 1.21 & 824.25 & -4.56 & -1.63 & 4.84 \\
2922522148821652224 & 104.8675 & -23.1750 & 234.5954 & -8.6721 & 1.20 & 830.94 & -4.48 & -1.31 & 4.67 \\
2922468100954457984 & 104.8751 & -23.4187 & 234.8204 & -8.7722 & 1.18 & 846.89 & -4.49 & -1.30 & 4.68 \\
2922345849002149632 & 103.7801 & -24.1939 & 235.0901 & -10.0106 & 1.18 & 848.63 & -4.41 & -1.13 & 4.55 \\
2921538768803796096 & 104.6908 & -24.5209 & 235.7518 & -9.4024 & 1.16 & 860.59 & -4.53 & -1.23 & 4.70 \\
2922437245905405056 & 104.4736 & -23.6385 & 234.8588 & -9.1992 & 1.13 & 885.32 & -4.33 & -0.93 & 4.43 \\
\hline \hline \\
Subgroup 3 (Green) & & & &  &  &  &  &  & \\
5610893151774163968 & 105.3863 & -26.4133 & 237.7574 & -9.6541 & 1.30 & 771.13 & -4.55 & -1.25 & 4.72 \\
5610899375187534464 & 105.6674 & -26.2780 & 237.7444 & -9.3685 & 1.27 & 789.09 & -4.55 & -1.46 & 4.77 \\
2921169096683353344 & 105.3423 & -25.5709 & 236.9701 & -9.3252 & 1.25 & 800.83 & -4.45 & -1.42 & 4.67 \\
5610889067266162944 & 105.4676 & -26.4752 & 237.8459 & -9.6153 & 1.25 & 800.86 & -4.56 & -1.46 & 4.79 \\
5611293550106938624 & 105.6130 & -25.9966 & 237.4659 & -9.2903 & 1.23 & 809.77 & -4.54 & -1.22 & 4.70 \\
5610840620033739136 & 105.5146 & -26.5679 & 237.9491 & -9.6174 & 1.23 & 810.13 & -4.48 & -1.42 & 4.70 \\
5610823715044523392 & 105.8065 & -26.7155 & 238.1986 & -9.4462 & 1.23 & 813.50 & -4.55 & -1.35 & 4.75 \\
5610686825844455808 & 105.2845 & -27.2988 & 238.5289 & -10.1174 & 1.23 & 814.10 & -4.43 & -1.17 & 4.59 \\
5610921842155684736 & 105.4806 & -25.9853 & 237.4033 & -9.3926 & 1.22 & 819.62 & -4.42 & -1.18 & 4.57 \\
2921232421679910144 & 105.1757 & -25.2661 & 236.6257 & -9.3286 & 1.22 & 819.83 & -4.51 & -1.16 & 4.66 \\
5610842230641908224 & 105.5902 & -26.5317 & 237.9457 & -9.5408 & 1.15 & 872.45 & -4.68 & -1.19 & 4.83 \\
\hline \hline \\
Subgroup 4 (Magenta) & & & &  &  &  &  &  & \\
5609602535580397440 & 105.9793 & -28.5000 & 239.8987 & -10.0792 & 1.29 & 773.82 & -4.63 & -1.68 & 4.93 \\
5610398513279813760 & 105.9350 & -28.0656 & 239.4835 & -9.9273 & 1.28 & 781.16 & -4.64 & -1.62 & 4.91 \\
5610509254716733824 & 106.3458 & -27.7197 & 239.3268 & -9.4502 & 1.22 & 821.44 & -4.49 & -1.47 & 4.72 \\
5610429024727430144 & 105.7851 & -27.6704 & 239.0636 & -9.8763 & 1.16 & 860.00 & -4.41 & -1.37 & 4.62 \\
5609574875992081280 & 106.4125 & -28.5572 & 240.1178 & -9.7606 & 1.16 & 860.73 & -4.79 & -1.71 & 5.09 \\
5610414215682818432 & 105.7387 & -28.0229 & 239.3686 & -10.0652 & 1.16 & 861.47 & -4.49 & -1.22 & 4.65 \\
5609585012113549056 & 106.5067 & -28.3212 & 239.9387 & -9.5837 & 1.15 & 868.30 & -4.38 & -1.29 & 4.57 \\
5609651326408739968 & 106.1222 & -28.1679 & 239.6496 & -9.8224 & 1.13 & 882.36 & -4.41 & -1.31 & 4.60 \\
\hline
\end{tabular}
\end{table*}


\bsp	
\label{lastpage}
\end{document}